\documentclass[aps,prb,twocolumn,amsmath,amssymb,superscriptaddress,citeautoscript,longbibliography,floatfix]{revtex4-1}

\usepackage{graphicx}% Include figure files
\usepackage{dcolumn}% Align table columns on decimal point
\usepackage{bm}
\usepackage{color}
\usepackage{mathptmx}
\usepackage{mathrsfs}
\usepackage{multirow}
\usepackage{natbib}

\begin{document}

\title{Strain and order-parameter coupling in Ni-Mn-Ga Heusler alloys \\ from resonant ultrasound spectroscopy}

\author{C.~Salazar Mej\'{i}a}\email{c.salazar-mejia@hzdr.de}
\affiliation{Max Planck Institute for Chemical Physics of Solids, N\"{o}thnitzer Str. 40, 01187 Dresden, Germany}
\affiliation{Present Address: Dresden High Magnetic Field Laboratory (HLD-EMFL), Helmholtz-Zentrum Dresden-Rossendorf, 01328 Dresden, Germany}

\author{N.~-O.~Born}
\affiliation{Max Planck Institute for Chemical Physics of Solids,  N\"{o}thnitzer Str. 40, 01187 Dresden, Germany}

\author{J.~A.~Schiemer}
\affiliation{Department of Earth Sciences, University of Cambridge, Downing Street, Cambridge CB2 3EQ, UK}

\author{C.~Felser}
\affiliation{Max Planck Institute for Chemical Physics of Solids,  N\"{o}thnitzer Str. 40, 01187 Dresden, Germany}

\author{M.~A.~Carpenter}
\affiliation{Department of Earth Sciences, University of Cambridge, Downing Street, Cambridge CB2 3EQ, UK}

\author{M.~Nicklas}\email{Michael.Nicklas@cpfs.mpg.de}
\affiliation{Max Planck Institute for Chemical Physics of Solids,  N\"{o}thnitzer Str. 40, 01187 Dresden, Germany}

\date{\today}

%%%%%%%%%%%%%%% Abstract%%%%%%%%%%%%%%%%
\begin{abstract}

Resonant ultrasound spectroscopy and magnetic susceptibility experiments have been used to characterize strain coupling phenomena associated with structural and magnetic properties of the shape-memory Heusler alloy series Ni$_{50+x}$Mn$_{25-x}$Ga$_{25}$ ($x=0$, 2.5, 5.0, and 7.5). All samples exhibit a martensitic transformation at temperature $T_M$ and ferromagnetic ordering at temperature $T_C$, while the pure end member ($x=0$) also has a premartensitic transition at $T_{PM}$, giving four different scenarios: $T_C>T_{PM}>T_M$, $T_C>T_M$ without premartensitic transition, $T_C\approx T_M$, and $T_C<T_M$.
Fundamental differences in elastic properties \emph{i.e.}, stiffening versus softening, are explained in terms of coupling of shear strains with three discrete order parameters relating to magnetic ordering, a soft mode and the electronic instability responsible for the large strains typical of martensitic transitions. Linear-quadratic or biquadratic coupling between these order parameters, either directly or indirectly via the common strains, is then used to explain the stabilities of the different structures. Acoustic losses are attributed to critical slowing down at the premartensite transition, to the mobility of interphases between coexisting phases at the martensitic transition and to mobility of some aspect of the twin walls under applied stress down to the lowest temperatures at which measurements were made.

\end{abstract}

\maketitle

%%%%%%%%%%%%%%%% Introduction %%%%%%%%%%%%%%%%
\section{INTRODUCTION}

Ni-Mn-Ga alloys that undergo a martensitic transformation, exhibit a shape-memory effect, specifically, a magnetic field induced structural reorientation. Large deformation values up to 12\% in low magnetic fields have been reported.\cite{Sozinov2013}  Shape-memory materials, more generally, can be used in applications including as actuators and sensors or for energy harvesting.\cite{Heczko2014} Additionally, the materials of this family can exhibit a giant magnetocaloric effect.\cite{Pareti_EPJB_2003} The study of the lattice dynamics which underpin the martensitic transformations and their characteristic strain behavior is important for a deeper understanding of the physical mechanisms behind the multifunctional properties of these materials. Resonant ultrasound spectroscopy (RUS) gives information on the coupling of the order parameter with strain but also on relaxation phenomena.\cite{SalazarMejia2015} Previous studies with RUS, or similar techniques, on different Ni-Mn-Ga samples have focused on the premartensitic transition, the damping properties and the determination of the elastic constants of the alloys.\cite{Sanchez-Alarcos2006,Perez-Landazabal2007,Chernenko2002,Chang2008,Aaltio2008,Seiner2009,Sedlak2017} RUS experiments in magnetic field have been used to evaluate magneto-elastic properties in Ni$_2$MnGa.\cite{Heczko2013,Seiner2013a,Seiner2014} Nevertheless, a combined systematic RUS and magnetic susceptibility study investigating the magneto-structural properties at the ferromagnetic, martensitic and premartensitic transitions is still missing.

\begin{figure}[b!]
\begin{center}
  \includegraphics[width=1.0\linewidth]{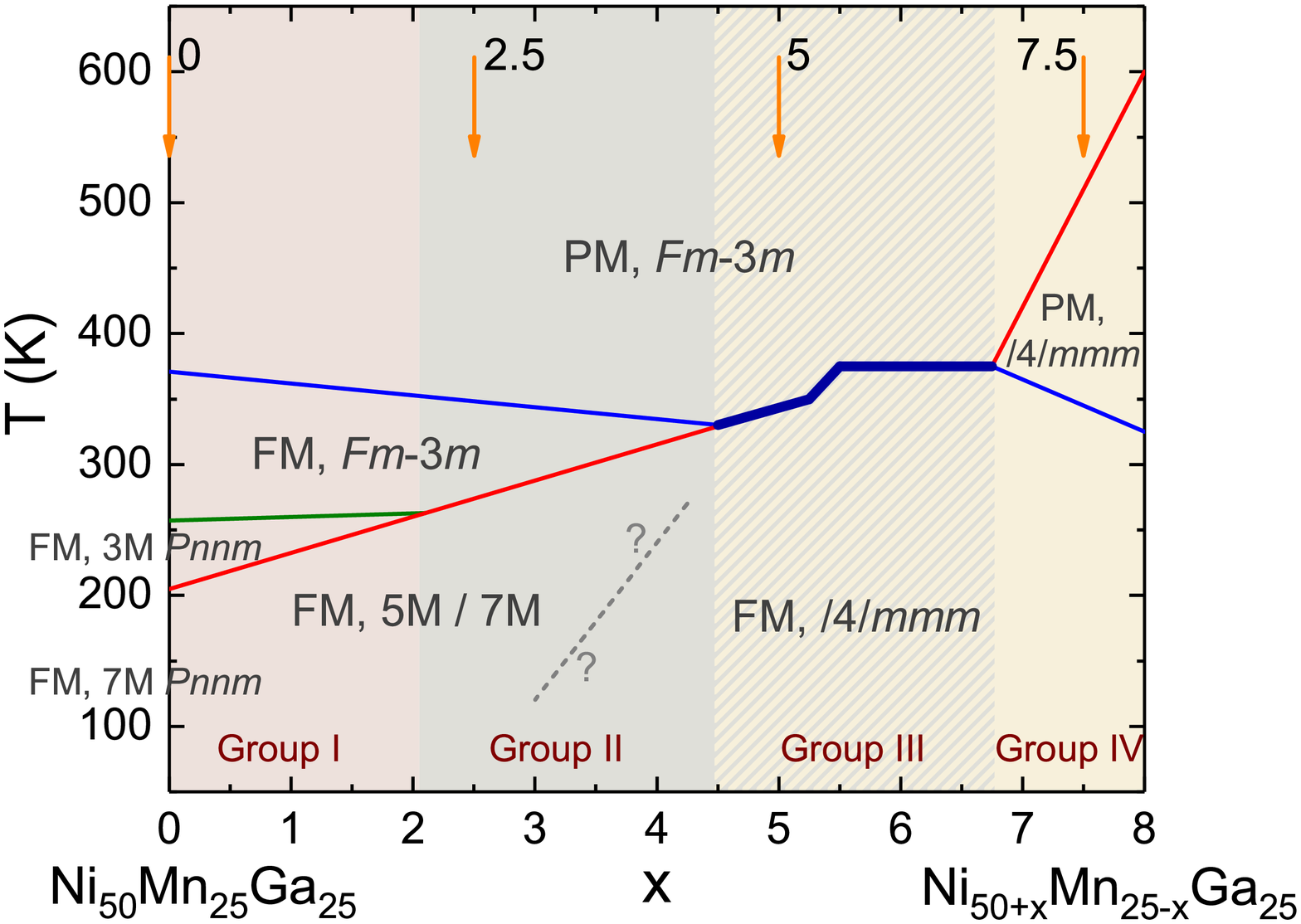}
  \caption{Schematic phase diagram for the Ni-rich end of the system Ni$_{50+x}$Mn$_{25-x}$Ga$_{25}$ (following Ref.\ \onlinecite{Vasilev1999,Vasilev2003,Khovaylo2005,Entel2014}). The boundary between stability fields of 5M/7M martensites and the nonmodulated, $I4/mmm$ structure has not yet been determined. Ferromagnetic structures become stable below the blue line ($T_C$) and martensitic structures become stable below the red line ($T_M$). The green line is $T_{PM}$, which marks the transition from the ferromagnetic austenite structure to the premartensite 3M structure.}\label{phaseDiag}
\end{center}
\end{figure}

In this work, we have studied polycrystalline Ni$_{50+x}$Mn$_{25-x}$Ga$_{25}$ Heusler alloys with $x=0$, 2.5, 5.0, and 7.5 using RUS and magnetic susceptibility measurements to focus on the temperature range which includes the ferromagnetic and the martensitic phase transitions. The selected concentrations belong to characteristically different groups in the Ni$_{50+x}$Mn$_{25-x}$Ga$_{25}$ phase diagram displayed in Fig.\ \ref{phaseDiag}.\cite{Chernenko1995,Chernenko1999,Chernenko2002,Vasilev1999,Vasilev2003,Khovaylo2005,Entel2014} The sample with $x=0$ belongs to group I, in which the ferromagnetic transition occurs at values of $T_C$ that are substantially above the martensitic transition temperature $T_M$ and in which there is an intermediate field of stability for the premartensite structure, $T_C > T_{PM} > T_M$. The sample with $x = 2.5$ belongs to group II, in which $T_M$ is close to room temperature, still below $T_C$ but without an intermediate phase. The sample with $x = 7.5$ belongs to group IV, which has $T_M > T_C$. The sample with $x = 5.0$ falls in group III, in that $T_C$ is expected to be more or less coincident with $T_M$. We have observed fundamental differences in the behavior of the crystal lattice at the martensitic transition, \emph{i.e.}, stiffening versus softening, determined by the presence or absence of a premartensitic transition and the relation between the Curie temperature and the martensitic-transformation temperature. The elastic and anelastic anomalies reveal the form and strength of coupling of strain with three separate order parameters which combine to give the soft mode, martensitic transition and ferromagnetic ordering observed in Heusler alloys.

\begin{figure}[b!]
\begin{center}
 \includegraphics[width=1.0\linewidth]{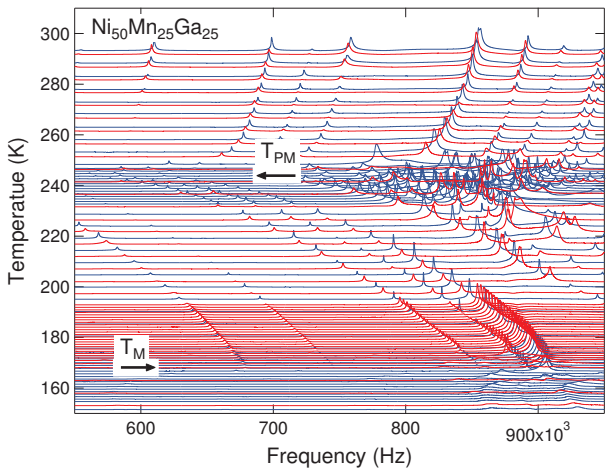}
  \caption{Segments of RUS spectra for Ni$_{50}$Mn$_{25}$Ga$_{25}$. The ordinate is the amplitude of the RUS spectra in arbitrary units. Each spectrum has been offset in proportion to the temperature at which it was collected. Accordingly, the axis is labeled as temperature. Blue traces are spectra collected during cooling and red traces are spectra collected during heating.}\label{spectra}
\end{center}
\end{figure}

%%%%%%%%%%%%%%% Experimental %%%%%%%%%%%%%%%%%%%
\section{EXPERIMENTAL}

The polycrystalline samples of Ni$_{50+x}$Mn$_{25-x}$Ga$_{25}$,  with a nominal $x=0$, 2.5, 5.0, and 7.5, were prepared by arc-melting the stoichiometric amounts of the elements under argon atmosphere. The ingots were then enclosed in tantalum ampules under argon atmosphere and, subsequently, sealed in evacuated quartz ampules and homogenized at 1073~K for 3 weeks. Afterwards they were quenched in cold water. The high quality of the samples was confirmed by x-ray powder diffraction. To ensure homogeneity and confirm the chemical composition, wavelength dispersive x-ray analysis and light microscopy were used. The actual compositions of the samples have been determined to be Ni$_{50}$Mn$_{25.5}$Ga$_{24.5}$, Ni$_{52.5}$Mn$_{23.25}$Ga$_{24.25}$, Ni$_{55}$Mn$_{20.25}$Ga$_{24.75}$ and Ni$_{57.5}$Mn$_{18.25}$Ga$_{24.25}$ for the nominal concentrations $x=0$, 2.5, 5.0, and 7.5, respectively. In the following we will denote the samples by their nominal concentrations. Magnetization experiments were carried out in a magnetic properties measurement system (MPMS, Quantum Design). For the RUS investigations, all samples were cut in the form of approximately rectangular parallelepipeds with edge dimensions between $1$ and $4$~mm. These had masses between 15.1 and 113.6\,mg. RUS data were obtained using two different in-house built systems. In the high-temperature instrument, the sample sits lightly between the tips of alumina rods which protrude into a horizontal Netzsch $1600^\circ$C resistance furnace. The piezoelectric transducers are at the other end of the rods, outside the furnace.\cite{McKnight2008} In the low-temperature instrument the sample sits directly between the transducers and is suspended in an atmosphere of a few mbars of helium gas, within a helium-flow cryostat.\cite{McKnight2007} The spectra were collected in the frequency range between 50 and 1200~kHz upon cooling followed by heating in the low temperature instrument and \textit{vice versa} in the high-temperature instrument. At each set point a period of 20 minutes was allowed for thermal equilibration before the spectrum was collected. The frequency $f$ and width at half maximum $\Delta f$ of selected resonance peaks in the primary spectra were fit with an asymmetric Lorentzian function. In general, for a polycrystalline sample, $f^2$ of each peak scales with some combination of the shear and bulk moduli but, since the resonance modes involve predominantly shearing motions, the temperature variation of $f^2$ effectively reflects that of the shear modulus. The inverse mechanical quality factor $Q^{-1} =\Delta f/f$ is a measure of acoustic attenuation.

%%%%%%%%%%%%%%Results%%%%%%%%%%%%%%%%%%%%%%%%%%%
\section{RESULTS}

\begin{figure}[t!]
\begin{center}
  \includegraphics[width=1.0\linewidth]{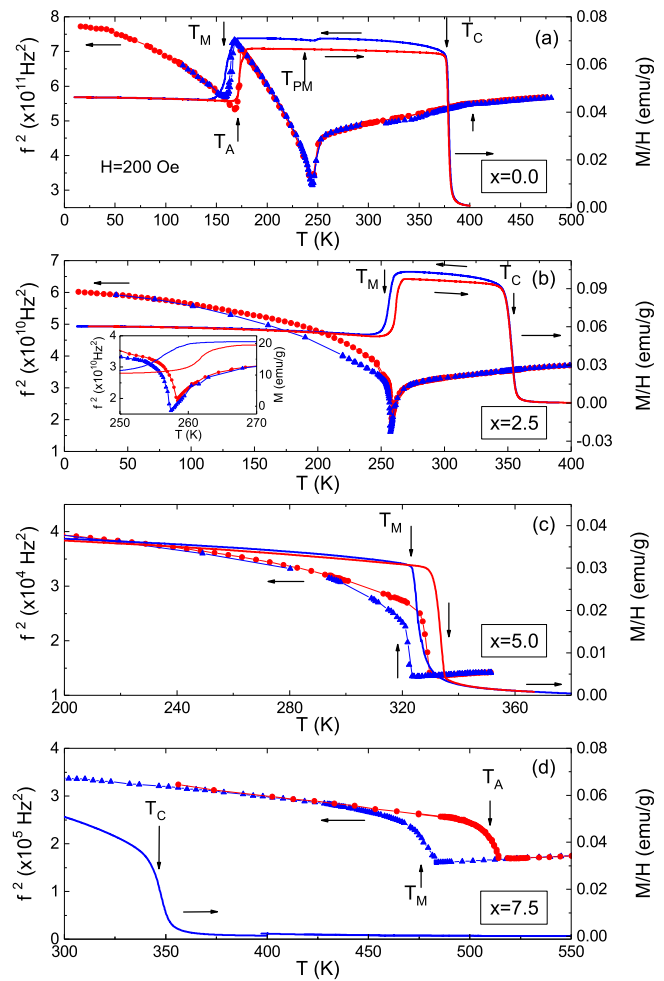}
  \caption{Temperature dependence $f^2$ (left axis) and magnetic susceptibility $M/H$ (right axis) for the series Ni$_{50+x}$Mn$_{25-x}$Ga$_{25}$ with (a) $x=0$, (b) $2.5$, (c) $5.0$, and (d) $7.5$. Blue triangles (lines) indicate data recorded on cooling and red circles (lines) on heating. Inset of (b) and (d) show the martensitic transition in detail. Magnetization measurements were performed at 200~Oe. }\label{RUS_all}
\end{center}
\end{figure}

%%%%%%%%%%%%%% RUS spectra%%%%%%%%%%%%%%%%%%%%%%%%%
Figure \ref{spectra} presents segments of RUS spectra for Ni$_{50}$Mn$_{25}$Ga$_{25}$ ($x=0$) in the low temperature range, recorded during both cooling and heating. Ni$_{50}$Mn$_{25}$Ga$_{25}$ orders ferromagnetically at $T_C = 380$~K. On cooling it undergoes a martensitic transformation at $T_M = 170$~K, which has a hysteresis of about 10~K. Additionally, this sample exhibits a premartensitic transition at $T_{PM} = 246$~K. The dependence of the resonant peaks with temperature can be inferred from the spectra. On cooling from the highest temperatures, a strong shift of the peaks toward lower frequencies (elastic softening) is observed as the temperature approaches the premartensitic transition at $T_{PM}$. Below this temperature, the peaks shift toward higher frequencies (stiffening) until reaching the martensitic transition at $T_M$. A marked increase in the width of the peaks and a decrease in the intensity is also evident below $T_M$. Above $T_M$ no thermal hysteresis is observed. The temperature dependence of $f^2$ and of the magnetic susceptibility $M/H$ for Ni$_{50+x}$Mn$_{25-x}$Ga$_{25}$ ($x=0$, 2.5, 5.0, and 7.5) are shown in Fig.\ \ref{RUS_all}. The $f^2(T)$ data have been obtained by fitting different resonant peaks, which have been combined by scaling to $f\approx0.7$, 0.18, 0.18, and 0.58~MHz at room temperature for $x=0$, 2.5, 5.0, and 7.5, respectively. In general, well defined peaks have been chosen, in order to be able to trace them in the whole temperature range. Furthermore, peaks at different resonant frequencies have been analyzed to confirm that the shear modulus is frequency independent. Magnetic susceptibility curves were recorded at 200~Oe under cooling and heating protocols. In the following we will discuss the experimental results in detail.

%%%%%%%%%%%%%%%%%%%%%%%%%% x=0 %%%%%%%%%%%%%%%%%%%%%%%%%%%%%%%%%%
The magnetic susceptibility of Ni$_{50}$Mn$_{25}$Ga$_{25}$ ($x=0$), displayed in Fig.\ \ref{RUS_all}a, shows three pronounced anomalies. On cooling, the magnetic susceptibility shows an increase at $T_C=380$~K indicating the ferromagnetic ordering. The pronounced drop in the magnetic susceptibility around $T_M=170$~K is the result of the martensitic transition at which the crystal structure of the sample changes from cubic to an incommensurate modulated structure.\cite{Singh2014} This first-order magneto-structural transition has a hysteresis between cooling and heating cycles ($T_A=180$~K) of around 10~K. Additionally, a dip at $T_{PM}= 246$~K is observed which corresponds to the premartensitic transition.\cite{Stuhr1997,Moya2006,Moya2009} These transitions are reflected in the elastic properties of the material. On cooling from high temperatures, softening of the shear modulus is evident on approaching $T_{PM}$, with an increase in slope below $T_C$. A strong dip in $f^2(T)$ is present at $T_{PM}$, as previously reported,\cite{Perez-Landazabal2007,Liu2013} without thermal hysteresis. Upon further cooling, a trend of stiffening is observed, followed by a step-like softening of about 22\% at $T_M$.

%%%%%%%%%%%%%%%%%%%%%%%%% x=2.5 %%%%%%%%%%%%%%%%%%%%%%%%%%%%%%%%%%%%
The dominant effect of Ni substitution on the Mn sites in the Ni$_{50+x}$Mn$_{25-x}$Ga$_{25}$ series is a shift of the martensitic transition toward higher temperatures and a shift of the ferromagnetic transition toward lower temperatures.\cite{Vasilev1999} Thus, the magnetic susceptibility in Ni$_{52.5}$Mn$_{22.5}$Ga$_{25}$ ($x=2.5$) displays similar behavior to Ni$_{50}$Mn$_{25}$Ga$_{25}$ (see Fig.\ \ref{RUS_all}b). The transitions move closer together, with $T_C$ decreasing to $353$~K and $T_{M}$ increasing to $256$~K, but the premartensitic transition is no longer observed. The elastic properties still show a dip in $f^2(T)$, but this is now located at $T_{M,A}$. Softening is observed as $T\rightarrow T_M$ from above, with a slight increase in slope at $T_C$. This becomes significantly steeper within $10$~K of $T_M$ (inset of Fig.\ \ref{RUS_all}b) and reverts to strong stiffening below $T_M$.

\begin{figure}[t!]
\begin{center}
  \includegraphics[width=0.95\linewidth]{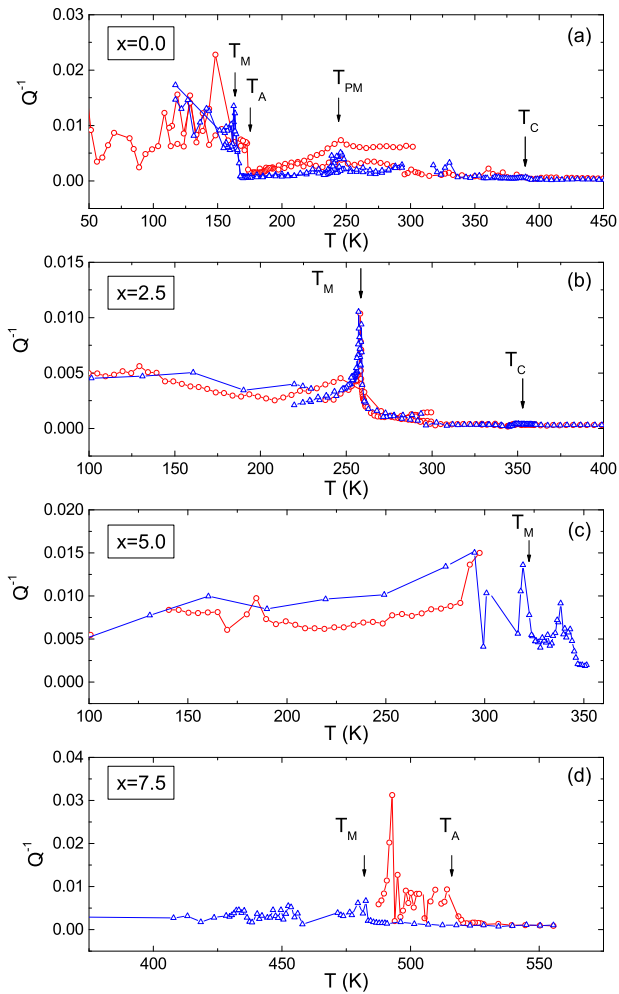}
  \caption{Temperature dependence of acoustic loss for the samples Ni$_{50+x}$Mn$_{25-x}$Ga$_{25}$ with (a) $x=0$, (b) $x=2.5$, (c) $x=5.0$ and (d) $x=7.5$. Resonant peaks in the primary spectra remain broad and weak at $T<T_M$, in comparison with those at $T>T_M$. Blue triangles indicate data recorded on cooling and red circles on heating.}\label{InvQ}
\end{center}
\end{figure}

%%%%%%%%%%%%%%%%%%%%%%%%%%% x=5.0 %%%%%%%%%%%%%%%%%%%%%%%%%%%%%%%
Upon further increasing the Ni content, the magnetic susceptibility curves change characteristically. The sample with $x=5.0$ displays only one transition in $M(T)/H$ (Fig.\ \ref{RUS_all}c). Upon cooling we find a step-like increase in $M(T)/H$ at 325~K, indicating ferromagnetic ordering. At the same time a hysteresis between heating and cooling cycles marks the first-order character expected for the martensitic transformation. We conclude that in Ni$_{55}$Mn$_{20}$Ga$_{25}$ the ferromagnetic and  martensitic transitions occur at the same temperature $T_M\approx T_C=325$~K. \emph{i.e.}, a transition from a paramagnetic-austenitic phase to a ferromagnetic-martensitic phase. The temperature dependence of $f^2$ exhibits a different pattern than observed for the previous samples with smaller $x$. Specifically, there is no dip in $f^2(T)$ throughout the whole temperature range. On cooling, a step-like stiffening of the lattice of about 77\% is detected at $T_M$. The stiffening is preceded by weak softening upon approaching $T_M$ from higher temperatures. We note that the slight difference in the transition temperatures determined from RUS and magnetization data might be accounted for by the applied field in case of the latter.

%%%%%%%%%%%%%%%%%%%%%%%%%%%% x=7.5 %%%%%%%%%%%%%%%%%%%%%%%%%%
In the sample with the highest Ni content of our investigation, $x=7.5$, the ferromagnetic transition occurs at $T_C=345$~K inside the martensitic phase, \textit{i.e.} $T_C<T_{M,A}$ (see Fig.\ \ref{RUS_all}d). The latter takes place at $T_M=476$~K ($T_A=510$~K) showing up in $f^2(T)$ as a pronounced stiffening of the lattice of about 145\%. The observed hysteresis of 34~K is larger than that observed in the samples with a smaller Ni concentration (ranging from 6 to 12~K). Around $T_{M,A}$ we also observe a small change in paramagnetic susceptibility $M(T)/H$. This reflects the reduction of the effective magnetic moments in the martensitic compared with the austenitic phase.\cite{Korolev2015,Vasilev1999} The overall temperature dependence of $f^2$ is similar to that of Ni$_{55}$Mn$_{20}$Ga$_{25}$, \emph{i.e.}, slight softening above $T_{M,A}$ followed by a step-like stiffening at the transition and weaker linear stiffening towards lower temperatures. There is no overt change in trend of $f^2$ at $T_C$.

%%%%%%%%%%%%%%%%%%% Acoustic loss %%%%%%%%%%%%%%%%%%%%%%%%%%%%
The temperature dependences of the inverse mechanical quality factor, $Q^{-1}$, representing acoustic loss, are plotted for all four samples in Fig.\ \ref{InvQ}. The transitions detected previously in $M(T)/H$ and $f^2(T)$ displayed in Fig.\ \ref{RUS_all} show corresponding anomalies in $Q^{-1}(T)$. Ni$_{50}$Mn$_{25}$Ga$_{25}$ exhibits a peak in $Q^{-1}(T)$ at $T_{PM}=246$~K and step-like changes at $T_{M}$ and $T_{A}$, respectively. There is, perhaps, a slight bump in the data for $Q^{-1}(T)$ at $T_C$ but it is substantially smaller than the clear peaks seen at lower temperatures. Ni$_{52.5}$Mn$_{22.5}$Ga$_{25}$ ($x=2.5$) exhibits a well pronounced peak in the acoustic loss at $T_{M,A}$. A possible slight anomaly at $T_C$ is again very small in comparison. The acoustic loss remained higher at all temperatures in the martensitic compared with the austenitic phase. Equivalent data for $x=5.0$  and $7.5$ show a single steep increase in acoustic loss at $T_M$, to the extent that it was not possible to measure peak widths at temperatures close to the transition point. Peaks in spectra collected from the martensitic phase then remained broad and weak.

%%%%%%%%%%%%%%%%%%%%%%%%%%%%%%%%%%%%%%DISCUSSION%%%%%%%%%%%%%%%%%%%%%%%%%%%%%%%%%%%%%%%%%%%
%%%%%%%%%%%%%%%%%%%%%%%%%%%%%%%%%%%%%%%%%%%

\section{DISCUSSION}

We note that all investigated samples exhibit a martensitic transformation from a cubic high-temperature austenitic phase to a low-symmetry martensitic phase and a ferromagnetic transition. However, there are substantial differences between the two transitions in the different samples that lead to distinct strain relaxation behavior. These differences are a reflection of the ways that three instabilities combine in Heusler compounds more generally. In the following, we address such differences from the perspective of how three discrete order parameters couple with strain and with each other, starting with an analysis of the fundamental constraints of symmetry.
%%%%%%%%%%%%

\subsection{Group theory}

Changes in elastic properties associated with phase transitions occur as a consequence of coupling of strain with the driving order parameter(s). In dynamic measurements, additional anelastic effects are typically due to fluctuations related to the order parameter(s) or of strain relaxation accompanying the motion of defects such as ferroelastic twin walls.

In the case of Ni-Mn-Ga alloys, there are three order parameters to consider, with symmetry properties that are taken here from a more comprehensive symmetry analysis of martensitic transitions.\cite{Carpenter2018} The ferromagnetic transition can be treated, in the simplest case, as having a single order parameter $Q_M$. The martensitic transitions are more complicated because they involve combinations of an electronic instability\cite{Brown1999,Brown2002} and a soft mode.\cite{Zheludev1995,Stuhr1997,Manosa2001} The order parameter for the electronic instability has the symmetry properties of the zone centre irreducible representation $\Gamma^+_3$ of parent space group $Fm\bar{3}m$. By itself this would give the nonmodulated (NM), tetragonal structure which has space group $I4/mmm$ (\emph{e.g.}\ Ref.\ \onlinecite{Banik2007}) and a single order parameter $Q_E$. The order parameter for the soft mode has symmetry properties related to points along the $\Sigma$ line between $\Gamma$ and N of the cubic Brillouin zone, \emph{i.e.}\ along a $\langle110\rangle^\ast$ direction. In the simplest case, this can also be expressed using a single order parameter, $Q_S$, conforming to $\Sigma_2$ symmetry, though it has $\Gamma^+_3$  and $\Gamma^+_5$ as secondary irreducible representation.

If $Q_S$ operates on its own, the resultant structure is the incommensurate premartensite which is commonly referred to as being the 3M structure. The commensurate equivalent would have space group $Pnnm$.\cite{Brown2002,Singh2015} The 7M martensite structure of Ni$_{50}$Mn$_{25}$Ga$_{25}$ arises as a consequence of coupling between $Q_E$ and $Q_S$. It has space group $Pnnm$,\cite{Brown2002,Ranjan2006} though it has also been proposed that the long repeat is incommensurate.\cite{Singh2014,Singh2015} A 5M structure can occur in Ni$_{50+x}$Mn$_{25-x}$Ga$_{25}$ when $x$ is small (\emph{e.g.}\ Refs.\ \onlinecite{Vasilev2003,Entel2014}) and the 5M and 7M structures have both been observed at $x = 4$.\cite{Khovaylo2004} Their stability limits have not yet been fully established.

\begin{table}[b!]
\begin{ruledtabular}
\small
\caption{Sequences of magnetic and structural states for the investigated samples of Ni$_{50+x}$Mn$_{25-x}$Ga$_{25}$ upon cooling.}
\begin{tabular}{c|c|c|>{$}c<{$}} %\hline
$x$ &  magnetic order   &  structure & Q  \\ \hline
\multirow{4}{*}
{0} & para & $Fm\bar{3}m$ & Q_M = Q_S = Q_E = 0 \\
  & ferro & $Fm\bar{3}m$ & Q_M \neq 0, Q_S = Q_E = 0 \\
  & ferro & 3M incommensurate & Q_M \neq 0, Q_S \neq 0, Q_E = 0 \\
	& ferro & 7M $Pnnm$ & Q_M \neq 0, Q_S \neq 0, Q_E \neq 0\\ \hline	
\multirow{3}{*}
{2.5} & para & $Fm\bar{3}m$ & Q_M = Q_S = Q_E = 0 \\
  & ferro & $Fm\bar{3}m$ & Q_M \neq 0, Q_S = 0, Q_E = 0 \\
  & ferro & 5M or 7M $Pnnm$ & Q_M \neq 0, Q_S \neq 0, Q_E \neq 0 \\ \hline	
\multirow{2}{*}
{5.0} & para & $Fm\bar{3}m$ & Q_M = Q_S = Q_E = 0 \\
  & ferro & $I4/mmm$  & Q_M \neq 0, Q_S = 0, Q_E \neq 0 \\  \hline
\multirow{3}{*}
{7.5} & para & $Fm\bar{3}m$ & Q_M = Q_S = Q_E = 0 \\
  & para & $I4/mmm$ & Q_M = 0, Q_S = 0, Q_E \neq 0 \\
  & ferro & $I4/mmm$ & Q_M \neq 0, Q_S = 0, Q_E \neq 0 \\  %\hline
\end{tabular}\label{trans}
\end{ruledtabular}
\end{table}

Order parameter combinations in structures which occur during heating and cooling of the four  Ni-Mn-Ga alloys used in the present study can be set out on the basis of the schematic phase diagram introduced in Fig.\ \ref{phaseDiag}. The sequences of magnetic and structural states and of the order parameters are summarized on Table\ \ref{trans}.

%%%%%%%%%%%%%%%%%%%%%%%%%%%%%%%%%
\subsection{Strain coupling}

The form of elastic softening or stiffening associated with each transition will depend on the form of coupling between individual strains and the three driving order parameters, the strength of coupling in each case and the thermodynamic character of the separate transitions. It has already been shown that the elastic properties of Ni-Mn-Ga alloys measured at frequencies of $\sim 1$~Hz show distinct general patterns of elastic softening/stiffening and acoustic loss, which reflect the different groups distinguished by their relative values of $T_C$ and $T_M$.\cite{Chernenko2002,Chang2008} Our present RUS data in combination with data from the literature allow the underlying causes of these to be set out more explicitly.

Acoustic resonances of a small sample of typical metal or ceramic are determined predominantly by shearing so that $f^2$ for each resonance mode scales effectively with the shear modulus. For a cubic crystal, this in turn depends on the single crystal elastic constants $C_{44}$ and $C_{11}-C_{12}$. For the orthorhombic crystals the shear modulus will depend on the related shear elastic constants and will show the influence, in particular, of changes in $C_{66}$ due to coupling of the order parameters with the shear strain $e_6$ (see Appendix \ref{estimation} for details), which arises from irreducible representation $\Gamma^+_5$, and in $C_{11}-C_{12}$ due to coupling with the tetragonal shear strain $e_t$, which arises from irreducible representation $\Gamma^+_3$.

The ferromagnetic transition at about $370$~K in group I and II alloys ($x = 0$ and $2.5$ in the present study), would be expected to give a break in crystallographic symmetry such that the lattice geometry becomes tetragonal due to coupling of the form $\lambda e_tQ_M^2$. This would be expected to give rise to a step like softening at $T_C$.\cite{Carpenter1998a}  % (as in Fig.\ 5g, h, i of Ref.\ \onlinecite{Carpenter1998a}).
This is not observed, however, because the coupling coefficient is sufficiently small that the crystals remain metrically cubic.\cite{Brown2002} The observed slight softening must arise from the next highest coupling terms $\lambda e_t^2Q_M^2$  and $\lambda e_4^2Q_M^2$  which will give softening or stiffening (depending on the sign of the coupling coefficient, $\lambda$) proportional to $Q_M^2$.

\begin{figure}[t!]
\begin{center}
  \includegraphics[width=1.0\linewidth]{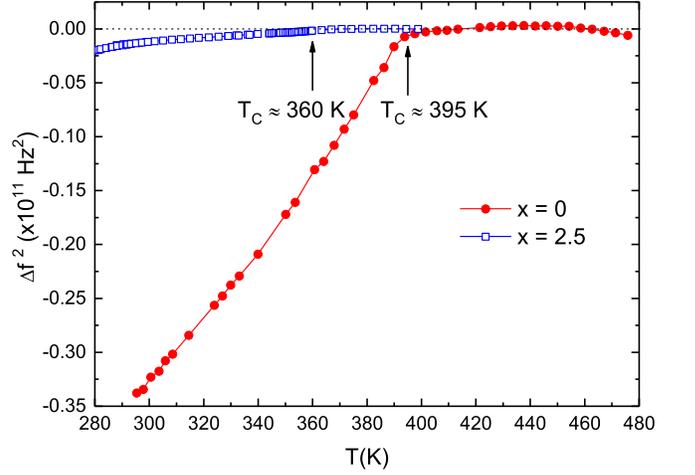}
  \caption{Variations of the change in $f^2(T)$, $\Delta f^2(T)$, with respect to a linear baseline fit to data in Fig.\ \ref{RUS_all} at $T > T_C$ for $x = 0$ and 2.5. $\Delta f^2(T)$ is expected to scale with the square of the ferromagnetic order parameter but its magnitude clearly reduces with increasing $x$.}\label{df2}
\end{center}
\end{figure}

Changes in the shear modulus, expressed as the difference $\Delta f^2$ between observed values of $f^2$ and a linear extrapolation of their values from above $T_C$, are shown in Fig.\ \ref{df2}. On this basis, the ferromagnetic transition is thermodynamically continuous and, as in the case of Ni$_{50}$Mn$_{35}$In$_{15}$,\cite{SalazarMejia2015} the effect is small.
RUS measurements on a single crystal of Ni$_{50}$Mn$_{25}$Ga$_{25}$ have previously shown that a significant contribution to the softening comes from $C_{11}-C_{12},$\cite{Seiner2009,Heczko2012,Seiner2013a,Seiner2013} but the contribution from $C_{44}$ is not yet known. The magnitude of softening is less at $x = 2.5$ than at $x = 0$ (see Fig.\ \ref{df2}), suggesting that the coupling coefficient for $\lambda e_t^2Q_M^2$ reduces with increasing $x$. Similar softening would be expected in association with the ferromagnetic transition in group IV alloys but there is no obvious deviation in $f^2$ below $T_C = 345$~K in data from the $x = 7.5$ sample (see Fig.\ \ref{RUS_all}d), implying that the strain coupling coefficients become negligibly small. An estimation of the magnitudes of symmetry breaking shear strains can be found in the Appendix \ref{estimation}.

%%%%%%%%%%%%%%%%%%%%%%%%%%%%%%%%%
\subsection{Elastic softening and stiffening}

$Q_E$ and $e_1-e_2$ have the symmetry properties of irreducible representation $\Gamma^+_3$  so that bilinear coupling of the form  $\lambda(e_1-e_2)Q_E$ is allowed. This gives rise to softening with falling temperature of $C_{11}-C_{12}$ as the martensitic transition is approached from above, typical of pseudoproper ferroelastic behavior. Single crystal measurements on stoichiometric Ni$_{50}$Mn$_{25}$Ga$_{25}$ at temperatures above $T_{PM}$ have shown this softening and, as expected because $e_6$ transforms as $\Gamma^+_5$  to give coupling of the form $\lambda e_6Q_E^2$, no equivalent softening was seen in $C_{44}$.\cite{Stipcich2004,Seiner2014} The softening seen in $f^2(T)$ for all four samples studied here arises from $C_{11}-C_{12}$, therefore.

The transition from austenite to 3M premartensite intervenes before the transition driven by $Q_E$ occurs. This is driven by the $\Sigma_2$ order parameter, $Q_S$, with which the strains from the irreducible representations $\Gamma^+_3$ and  $\Gamma^+_5$ couple as $\lambda(e_1-e_2)Q_S^2$  and $\lambda e_6Q_S^2$. No latent heat or structural discontinuity which would imply first order character for the transition has been yet observed,\cite{Zheludev1995,Kokorin1996,Manosa1997} and the heat capacity anomaly reported in Ref.\ \onlinecite{Manosa1997} for a crystal with nearly stoichiometric composition ($x = 0$) has a step at $T_{PM}$ consistent with the transition being second order. A small hysteresis has been reported in magnetic susceptibility and dynamical mechanical analysis (DMA) data for samples of Ni$_{50+x}$Mn$_{25-x}$Ga$_{25}$ with $0\leq x\leq 2$,\cite{Liu2013} but is probably accounted for by the effect of applied magnetic and stress fields. The transition is thus second order and improper ferroelastic, which is expected to give a stepwise softening of both $C_{11}-C_{12}$  and $C_{66}$.\cite{Carpenter1998a}

Essentially this pattern is seen in $C_{11}-C_{12}$ and $C_{44}$ from pulse-echo ultrasonic data for single crystals of Ni$_{50}$Mn$_{25}$Ga$_{25}$.\cite{Worgull1996,Stenger1998,Stipcich2004,Manosa1997} The frequency of the measurements was not always specified but is assumed to be around $10$~MHz. More generally, the elastic anomaly at $T_{PM}$ observed in measurements made at $\sim 1$~Hz\cite{Chernenko2002} and $10^5-10^6$~Hz, this study and Refs.\ \onlinecite{Sanchez-Alarcos2006,Perez-Landazabal2007,Seiner2013}, has steep softening in a temperature interval of up to $10$~K above the transition point followed by non-linear recovery below it. The precursor softening is typical of order parameter fluctuations ahead of improper ferroelastic and co-elastic transitions.\cite{Carpenter1998b,Carpenter2015} $C_{11}-C_{12}$ recovers to higher values than attained above $T_{PM}$, which is presumably a consequence of some contribution of bilinear coupling of $e_1-e_2$ with the  $\Gamma^+_3$ order parameter, once the cubic symmetry has been broken. The premartensite phase has a tweed microstructure, \cite{Zheludev1995,Chernenko2002} so that all these measurements are averages for crystals which may be orthorhombic only on a local length scale.\cite{Seiner2014}

There is no group/subgroup relationship between the premartensite structure and 5M/7M martensites. As a consequence, there is no order parameter which relates one directly to the other and the martensitic transition at small values of $x$ is necessarily first order. It is accompanied by a large increase in shear strain and the development of abundant ferroelastic twinning. The elastic anomalies are simply stepwise changes with significant hysteresis and little or no precursor effects,\cite{Sanchez-Alarcos2006,Perez-Landazabal2007} which is also seen in our data, $x = 0$ and 2.5 (group I and II compounds) presented in Figs.\ \ref{RUS_all}a and \ref{RUS_all}b. On the other hand, the austenite -- NM transition at larger values of $x$ (group III and IV alloys; $x = 5.0$ and 7.5 data shown in Figs.\ \ref{RUS_all}c and \ref{RUS_all}d) is driven by the  $\Gamma^+_3$ order parameter and is first order due to the presence of third order terms in the Landau expansion for excess free energy. Softening as $T_M$ is approached from above (see Fig.\ \ref{RUS_all}) becomes a steep increase in the shear modulus below $T_M$,\cite{Chang2008} as expected for pseudoproper ferroelastic character ($\lambda(e_1-e_2)Q_E$, $e_6 = 0$) when the transition is first order.\cite{Carpenter1998a} The same pattern has been observed at the austenite -- 5M/7M transition in Ni$_{50}$Mn$_{35}$In$_{15}$.\cite{SalazarMejia2015}

The premartensite and martensite transitions occur at about the same temperature when $x = 2.5$ ($T_{PM} \approx T_M$, see Fig.\ \ref{phaseDiag}). As a consequence, the variation of $f^2(T)$ displays aspects of both, with the precursor softening due to the proximity to $T_{PM}$ and the recovery below $T_M$ being that of the martensite.
%%%%%%%%%%%%%%%%%%%%%%%%%%%%%

\subsection{Acoustic loss}

The patterns of acoustic loss observed by RUS presented in Fig.\ \ref{InvQ} for the four different Ni-Mn-Ga samples are closely similar to those reported in the literature. In particular, transitions to the premartensitic 3M structure and to the NM martensite structure are marked by a sharp peak in attenuation associated with the shear modulus modes at the transition temperatures. The loss peak at $T_{PM}$ has been seen previously in measurements made at $0.1-5$~Hz by DMA,\cite{Chang2008,Liu2013} at $\sim1$~MHz by RUS,\cite{Sanchez-Alarcos2006,Perez-Landazabal2007,Seiner2013} and at $5-10$~MHz using pulse-echo ultrasonics.\cite{Worgull1996,Stenger1998} The same loss peak was also observed in inverted pendulum experiments.\cite{Gavriljuk2003} In Ni$_{50+x}$Mn$_{25-x}$Ga$_{25}$ with low values of $x$, at least, the transition is second order in character and the occurrence of the peak always at $T_{PM}$ suggests that the loss mechanism is due to critical slowing down of fluctuations in the order parameter $Q_S$.

The austenite -- NM martensite transition is first order so the loss mechanism is most likely related to the mobility under applied stress of interfaces between the transforming phases. A steep loss peak is typically observed also at the first order premartensite -- 5M/7M transition,\cite{Chang2008,Liu2013} which is again most likely due to mobility of interfaces between coexisting phases. In the case of Ni$_{50}$Mn$_{25}$Ga$_{25}$ the temperature interval of coexistence is about $40$~K,\cite{Singh2015a} which probably accounts for the broad loss peak seen in the present study.

More interesting is the observed loss behavior within the stability fields of the premartensite and martensite phases since this relates to the mobility of ferroelastic twin walls. Under the low stress conditions of an RUS experiment, $Q^{-1}$ remains relatively high below $T_M$ at compositions which fall within all four alloy groups,\cite{Carpenter2015} \emph{i.e.} for both 5M/7M and NM martensites, but reduces to low values in the stability field of the premartensite 3M phase.\cite{Sanchez-Alarcos2006,Perez-Landazabal2007,Seiner2013} This is also seen in our study.  The same has been reported in some,\cite{Chang2008} but not all DMA measurements,\cite{Liu2013} and in pulse-echo ultrasonic results.\cite{Stenger1998} The most likely explanation of the difference is that shear strains are also very much smaller in the premartensite structure than in the 5M/7M and NM martensites and that, as a consequence, the changes in strain state that occur when a twin wall is displaced are correspondingly much smaller. Alternatively, motion of individual twin walls might be jammed due to interactions between them in the tweed microstructure.

A second, broad loss peak has been seen in the stability fields of both 5M/7M martensite\cite{Aaltio2008,Chang2008} and premartensite\cite{Liu2013} as well as for the $x = 2.5$ sample in the present study, but these do not resemble the much steeper loss peaks associated with domain wall freezing seen, for example, in LaAlO$_3$.\cite{Harrison2004} The loss parameters remain high down to the lowest measuring temperatures, suggesting that at least some components of the twin walls remain mobile.

Possible mechanisms for a peak in acoustic loss at the ferromagnetic transition, such as at the antiferromagnetic ordering transition in CoF$_2$,\cite{Thomson2014} could include critical slowing down of fluctuations of the magnetic order parameter coupled with phonons. However, any increase in $Q^{-1}$ at $T_C$ of the polycrystalline samples with $x = 0$ and 2.5 appears to be negligibly small in the present study. Seiner \emph{et al.}\ also found no evidence for a loss peak in their RUS data from a slowly cooled single crystal of Ni$_{50}$Mn$_{25}$Ga$_{25}$,\cite{Seiner2013} but they found an increase in $Q^{-1}$ below $T_C$ when a complex microstructure of interacting magnetic domain walls and fine scale antiphase domains was induced by quenching from high temperatures. This reached a maximum at $T_{PM}$, suggesting that it was related to the premartensitic transition rather than simply to the ferromagnetic ordering.

\subsection{Order parameter coupling}
As well as each of the three order parameters coupling with strain, it is inevitable that they will couple with each other, either directly or via the common strain. Direct coupling terms allowed by symmetry include $\lambda Q_EQ_S^2$,  $\lambda Q_EQ_M^2$  and $\lambda Q_M^2Q_S^2$. Biquadratic coupling is always allowed and can lead to sequences of structural states involving only one order parameter or both in standard patterns.\cite{Salje1986} The consequences of linear-quadratic coupling have only been recently considered in general terms,\cite{Salje2011} and the predicted structural sequences match aspects of the relationships between structures in Fig.\ \ref{phaseDiag}. In particular, for $T_{PM} > T_M$, the expected sequence would be a second order transition to a state with $Q_S \neq 0$, $Q_E = 0$ followed by a first order transition to a state with $Q_S \neq 0$, $Q_E \neq 0$. For $T_{PM} < T_M$, a single phase transition to a state with $Q_S \neq 0$, $Q_E \neq 0$ is expected because $Q_E$ acts as a field for $Q_S$. This is exactly the change seen as $T_{PM}$ and $T_M$ converge with increasing $x$, and the same arguments apply to the convergence of $T_C$ and $T_M$.

%%%%%%%%%% Conclusions $$$$$$$$$$$$$$$$$$$$
\section{CONCLUSIONS}

Combined elasticity and magnetic susceptibility measurements from a set of representative samples belonging to the series Ni$_{50+x}$Mn$_{25-x}$Ga$_{25}$ ($x=0$, 2.5, 5.0, and 7.5) have provided a coherent picture of the consequences of strain coupling effects associated with the particular combination of instabilities that is commonly observed in Heusler alloys. These produce characteristic precursor softening of $C_{11}-C_{12}$ over a wide temperature interval due to bilinear coupling of $e_t$ with the electronic order parameter ($\lambda e_tQ_E$, pseudoproper ferroelastic behavior), stepwise softening below $T_M$ due to linear-quadratic coupling of both $e_t$ and $e_6$ with the order parameter for the structural transition driven by the soft mode ($\lambda e_tQ_S^2$, $\lambda e_6Q_S^2$, improper ferroelastic behavior) and biquadratic coupling of $e_t$ and $e_4$ with the ferromagnetic order parameter ($\lambda e_t^2Q_M^2$, $\lambda e_4^2Q_M^2$). The strength of coupling is very substantially greatest for coupling of $e_t$ with $Q_E$, giving rise to the large shear strains typical of martensitic phase transitions. Coupling of these three order parameters can account for the topology of the phase diagram, as well as for the particular structure types which are observed in Ni-Mn-Ga alloys. Under the low stress and high frequency conditions of an RUS experiments, acoustic losses occur below $T_M$ down to the lowest temperatures at which measurements were made. These have been assumed to relate to mobility of some component of the ferroelastic twin walls, without any indication that they become frozen or pinned.

\appendix
\section{SYMMETRY BREAKING SHEAR STRAINS}\label{estimation}

\begin{table*}[tbh!]
\begin{center}
\renewcommand{\arraystretch}{1.5}
\caption{Expressions used to estimate tetragonal and orthorhombic shear strains, $e_t$ and $e_6$, in terms of lattice parameters for commensurate 7M, 3M and NM structures. The 3M structure is actually incommensurate so, in practice, the commensurate repeat distance for $b$ is taken from the pseudocubic lattice parameters of the orthorhombic structure. Reference axes, $X$, $Y$ and $Z$ have been taken as parallel to the crystallographic axes of the parent cubic structure which has lattice parameter $a_0$.}
\begin{tabular}{|>{$}c<{$}|>{$}c<{$}|>{$}c<{$}|} \hline
\text{7M} &  \text{3M}    &  \text{NM}  \\ \hline \hline
e_1+e_2=\frac{\sqrt{2}a-a_o}{a_o}+\frac{\left(\sqrt{2}/7\right)b-a_o}{a_o} & e_1+e_2=\frac{\sqrt{2}a-a_o}{a_o}+\frac{\left(\sqrt{2}/3\right)b-a_o}{a_o} & e_1=e_2=\frac{\sqrt{2}a-a_o}{a_o}
\\ \hline
e_3=\frac{c-a_o}{a_o} & e_3=\frac{c-a_o}{a_o} & e_3=\frac{c-a_o}{a_o}\\ \hline
e_t=\frac{1}{\sqrt{3}}(2e_3-e_1-e_2) & e_t=\frac{1}{\sqrt{3}}(2e_3-e_1-e_2) & e_t=\frac{1}{\sqrt{3}}(2e_3-e_1-e_2)\\ \hline
e_6=\frac{\left(\sqrt{2}/7\right)b-a_o}{a_o}-\frac{\sqrt{2}a-a_o}{a_o} & e_6=\frac{\left(\sqrt{2}/3\right)b-a_o}{a_o}-\frac{\sqrt{2}a-a_o}{a_o} & e_6=0\\ \hline
a_\textit{o}\approx\left(\frac{2}{7}abc\right)^{1/3} & a_o\approx\left(\frac{2}{3}abc\right)^{1/3} & a_\textit{o}\approx(2abc)^{1/3}\\ \hline
\end{tabular}\label{strains}
\end{center}
\end{table*}

The magnitudes of symmetry breaking shear strains accompanying the premartensite and martensite transitions can be determined from lattice parameter data using the expressions set out in Table \ref{strains}. Representative values of $e_t$ and $e_6$ calculated from lattice parameters given by Ref.\ \onlinecite{Brown2002} for the 7M structure of Ni$_{50}$Mn$_{25}$Ga$_{25}$ at 20~K, are -0.076 and 0.007 respectively. The value of $e_t$ for a tetragonal nonmodulated structure calculated from the lattice parameters of Ni$_{54.5}$Mn$_{21.5}$Ga$_{24}$ at room temperature\cite{Gavriljuk2003} is 0.217. By way of contrast, the premartensite 3M structure has much smaller shear strains, $e_t = -0.007$, $e_6 = 0.002$, as calculated in a related manner using lattice parameters given by Ref.\ \onlinecite{Ohba2005} for Ni$_{50}$Mn$_{25}$Ga$_{25}$ at 250~K.

%%%% acknowledgments%%%
\begin{acknowledgments}
This work was financially supported  by the ERC Advanced Grant (291472) "Idea Heusler". RUS facilities in Cambridge have been supported by grants from the Natural Environment Research Council (NE/B505738/1, NE/F017081/1).and the Engineering and Physical Sciences Research Council (EP/I036079/1).
\end{acknowledgments}

\newpage

%%%%%%%%%% Bibliography %%%%%%%%%%%%%%%%%%%%%%%%%%%
\bibliography{References_bibtex}

\end{document}